\documentclass[11pt]{article}
\usepackage{amsmath,amssymb,amsthm,amsxtra,overpic,bbm,bm,epsfig,ulem,color,multirow}
\usepackage{epstopdf}
\usepackage{braket}

\begin{document}

\begin{center}
{\Large Rescuing flavor-symmetry-forbidden leptogenesis \\ within the left-right symmetric framework}
\end{center}

\vspace{0.05cm}

\begin{center}
{\bf Yan Shao, \bf Zhen-hua Zhao\footnote{Corresponding author: zhaozhenhua@lnnu.edu.cn}} \\
{ $^1$ School of Physics and Electronic Technology, Liaoning Normal University, Dalian 116029, China \\
$^2$ Center for Theoretical and Experimental High Energy Physics, \\ Liaoning Normal University, Dalian 116029, China }
\end{center}

\vspace{0.2cm}

\begin{abstract}
While the type-I seesaw model provides a unified framework for explaining the origin of neutrino masses and the baryon asymmetry of the Universe, flavor symmetries offer an attractive approach to understanding the observed neutrino mixing pattern. However, in many flavor-symmetry-based type-I seesaw models, either the Dirac neutrino mass matrix $M^{}_{\rm D}$ or the right-handed-neutrino mass matrix $M^{}_{\rm R}$ takes an exactly degenerate form, which prevents the conventional leptogenesis mechanism from working. In this paper, without breaking the original flavor structure dictated by the employed flavor symmetries, we investigate whether such forbidden leptogenesis scenarios can be rescued within the framework of the left-right symmetric model. We show that, in these scenarios, the contribution of the Higgs triplet present in the left-right symmetric model to the CP asymmetries of right-handed neutrino decays can successfully reproduce the observed baryon asymmetry. The leptogenesis mechanism proposed here can also be directly applied in more general type-I+II seesaw models.
\end{abstract}

\newpage

\section{Introduction}

The Standard Model (SM) of particle physics has achieved remarkable success in describing most experimental observations. Nevertheless, it fails to account for two fundamental phenomena: the origin of neutrino masses and the observed baryon asymmetry of the Universe (BAU). On the one hand, neutrino oscillation experiments have firmly established that neutrinos are massive and mixed states~\cite{xing}, in clear contradiction with the SM prediction of massless neutrinos. On the other hand, cosmological observations indicate a tiny but non-vanishing excess of baryons over antibaryons, quantified by~\cite{planck}
\begin{eqnarray}
Y^{}_{\rm B} \equiv \frac{n^{}_{\rm B}-n^{}_{\rm \bar B}}{s} \simeq (8.69 \pm 0.04)\times10^{-11}\;,
\label{1.1}
\end{eqnarray}
where $n^{}_{\rm B}$ ($n^{}_{\rm \bar B}$) denotes the baryon (antibaryon) number density and $s$ is the entropy density.

A particularly appealing framework capable of addressing both problems simultaneously is the type-I seesaw model~\cite{seesaw1}-\cite{seesaw5}, in which the SM is extended by three heavy right-handed neutrinos (RHNs) $N^{}_I$ ($I=1,2,3$). The Yukawa interactions between left- and right-handed neutrinos generate the Dirac mass matrix $M^{}_{\rm D}=Y^{}_\nu v$, where $Y^{}_\nu$ is the neutrino Yukawa coupling matrix and $v=174~{\rm GeV}$ is the Higgs vacuum expectation value (VEV), while the RHNs possess a Majorana mass matrix $M^{}_{\rm R}$. Under the seesaw condition $M^{}_{\rm R}\gg M^{}_{\rm D}$, the effective light-neutrino mass matrix is given by
\begin{eqnarray}
M^{}_{\nu}=-M^{}_{\rm D} M^{-1}_{\rm R} M^T_{\rm D}\;.
\label{1.2}
\end{eqnarray}
This framework naturally explains the smallness of neutrino masses, and simultaneously provides an elegant mechanism for generating the BAU through leptogenesis~\cite{leptogenesis}-\cite{Lreview4}.

The observed neutrino mixing pattern, established by neutrino oscillation experiments (see the global-fit results in Refs.~\cite{global1, global2}), suggests the possible existence of an underlying flavor structure in the lepton sector. And non-Abelian discrete flavor symmetries (such as the ${\rm A}^{}_4$ and ${\rm S}^{}_4$ groups) provide a particularly attractive approach to understanding these observations~\cite{FS1}-\cite{FS4}.
In many type-I seesaw models based on non-Abelian flavor symmetries, the three lepton doublets and the three right-handed neutrinos are organized into a triplet representation, respectively. In the symmetry limit, this assignment naturally leads to highly constrained neutrino mass matrices. To be specific, either the Dirac neutrino mass matrix or the RHN mass matrix is constrained into the following simple form ~\cite{FS1}-\cite{FS4}
\begin{eqnarray}
M^{}_{\rm D} = m^{}_{\rm D} \begin{pmatrix}
1 & 0 & 0 \\
0 & 0 & 1 \\
0 & 1 & 0
\end{pmatrix}
 \qquad \text{or} \qquad M^{}_{\rm R} = M_0
\begin{pmatrix}
1 & 0 & 0 \\
0 & 0 & 1 \\
0 & 1 & 0
\end{pmatrix} \;.
\label{1.3}
\end{eqnarray}
with $m^{}_{\rm D}$ and $M^{}_0$ denoting their overall scales. Such exactly degenerate structures naturally arise and commonly exist in many realistic type-I seesaw models based on non-Abelian flavor symmetries. Since the exact flavor-symmetry limit generally cannot reproduce the observed neutrino mixing pattern, the flavor symmetry must be broken in a suitable way. The simplest and most natural possibility is that only one of $M^{}_{\rm D}$ and $M^{}_{\rm R}$ retains the exactly degenerate forms in Eq.~(\ref{1.3}), while the other acquires a nontrivial flavor structure from certain breaking patterns of the employed flavor symmetries.

Despite their simplicity, the exactly degenerate forms of $M^{}_{\rm D}$ and $M^{}_{\rm R}$ in Eq.~(\ref{1.3}) prevent the conventional leptogenesis mechanism from working: for such a form of $M^{}_{\rm D}$, the flavor symmetry enforces orthogonality relations among different columns of the neutrino Yukawa coupling matrix, leading to vanishing CP asymmetries of RHN decays (see the discussion around Eq.~(\ref{2.5})); for such a form of $M^{}_{\rm R}$, three RHNs are exactly degenerate in mass, which also leads to vanishing CP asymmetries of RHN decays (see the discussion around Eq.~(\ref{3.1})). The interplay between flavor symmetries and leptogenesis has been extensively studied in the literature. For example, Ref.~\cite{Hagedorn:2017wjy} reviews an $A_4$ realization with the same leading-order Dirac neutrino mass texture as the $M_{\rm D}$ form in Eq.~(\ref{1.3}). In this realization, the conventional CP asymmetries vanish at leading order, whereas subleading symmetry breaking corrections to the Dirac neutrino mass matrix can make them non-vanishing. This provides a possible way to revive the conventional RHN-mediated leptogenesis through additional flavor symmetry breaking~\cite{Mohapatra:2004hta}--\cite{Mohapatra:2006se}.

In this paper, we explore alternative mechanisms that can rescue leptogenesis for the exactly degenerate forms of $M^{}_{\rm D}$ and $M^{}_{\rm R}$ in Eq.~(\ref{1.3}), without breaking the original flavor structure dictated by the employed flavor symmetries. We find that the left-right symmetric model (LRSM)~\cite{Pati:1974yy}-\cite{Senjanovic:1975rk} provides a particularly appealing framework for this purpose. As is well known, the LRSM is a well-motivated extension of the SM based on the gauge group ${\rm SU}(2)^{}_{\rm L} \times {\rm SU}(2)^{}_{\rm R} \times {\rm U}(1)^{}_{\rm B-L}$. In this framework, RHNs emerge as natural ingredients rather than ad hoc additions. Moreover, the LRSM contains two Higgs triplets $\Delta^{}_{\rm L,R}$. After they acquire non-zero VEVs $v^{}_{\rm L}$ and $v^{}_{\rm R}$, their Yukawa interactions generate the left- and right-handed neutrino mass matrices
\begin{eqnarray}
M^{}_{\rm L} = 2 v^{}_{\rm L} f \;, \qquad
M^{}_{\rm R} = 2 v^{}_{\rm R} f \;,
\label{1.4}
\end{eqnarray}
where $f$ denotes the Yukawa coupling matrix related to the Higgs triplets.
Consequently, the light-neutrino mass matrix receives simultaneous contributions from both type-I and type-II seesaw mechanisms
\begin{eqnarray}
M^{}_\nu
=
M^{\rm II}_\nu+M^{\rm I}_\nu
=
r M^{}_{\rm R}
-
M^{}_{\rm D} M^{-1}_{\rm R} M^{T}_{\rm D} \; ,
\label{1.5}
\end{eqnarray}
with $r  \equiv v^{}_{\rm L}/v^{}_{\rm R}$.
Remarkably, the LRSM introduces a new source of CP asymmetry of RHN decays that is absent in the conventional type-I seesaw framework: the Higgs triplet $\Delta^{}_{\rm L}$ generates additional contributions to the CP asymmetry of RHN decays (see the discussion around Eq.~(\ref{2.6}))~\cite{Hambye:2003ka}-\cite{Hambye}.
Motivated by this fact, we investigate whether the above-mentioned flavor-symmetry-forbidden leptogenesis can be rescued within the LRSM (specifically, via the contribution of $\Delta^{}_{\rm L}$ to the CP asymmetry for RHN decays) \footnote{For studies on leptogenesis in the LRSM with different focuses, see e.g., Refs.~\cite{LRSMlepto1}-\cite{LRSMlepto12}.}.
This scalar-triplet contribution also arises in more general type-I+II seesaw models, in which the SM is extended by an $SU(2)_{\rm L}$ Higgs triplet in addition to RHNs, without requiring left-right gauge symmetry. Its applicability beyond the LRSM will be discussed in Section~4.

The remaining parts of this paper are organized as follows. In Sections~2 and 3, we separately investigate the realization of leptogenesis within the LRSM for the forms of $M^{}_{\rm D}$ and $M^{}_{\rm R}$ in Eq.~(\ref{1.3}), respectively. In Section~4, we discuss the applicability of the scenarios considered in Sections~2 and 3 to more general type-I+II seesaw models. Finally, a summary of our main results is presented in Section~5.

\section{Leptogenesis for the exactly degenerate form of $M_{\rm D}$}

In this section, we investigate the realization of leptogenesis within the LRSM for the exactly degenerate form of $M^{}_{\rm D}$ in Eq.~(\ref{1.3}), while $M^{}_{\rm R}$ acquires a nontrivial flavor structure from certain symmetry-breaking patterns of the employed flavor symmetries. For illustration and definiteness, we take $M^{}_{\rm R}$ to have the texture that the resulting neutrino mixing pattern is of the TM1 type~\cite{TM1-1}-\cite{TM1-4}, which remains fully consistent with current neutrino oscillation data. In Section~2.1, we first introduce the model framework under consideration and some relevant leptogenesis basis. In Section~2.2, we then perform the leptogenesis calculations.

\subsection{Model framework and leptogenesis basis}

Historically, the facts that the neutrino mixing angles $\theta^{}_{12}$ and $\theta^{}_{23}$ are close to some special values (i.e., $\sin^2 \theta^{}_{12} \sim 1/3$ and $\sin^2 \theta^{}_{23} \sim 1/2$) and the smallness of $\theta^{}_{13}$ suggest that the neutrino mixing matrix approximates to a very special form (which is referred to as the tribimaximal (TBM) mixing \cite{TB1,TB2}) as
\begin{eqnarray}
U^{}_{\rm TBM}= \displaystyle \frac{1}{\sqrt 6} \left( \begin{array}{ccc}
-2 & \sqrt{2} & 0 \cr
1 &  \sqrt{2}  & -\sqrt{3}  \cr
1 &  \sqrt{2}  & \sqrt{3} \cr
\end{array} \right)  \;,
\label{2.1}
\end{eqnarray}
which naturally predicts $\sin^2 \theta^{}_{12} = 1/3$, $\sin^2 \theta^{}_{23} = 1/2$ and $\theta^{}_{13} =0$.
However, the observation of a sizable $\theta^{}_{13}$ requires modifications to the TBM mixing pattern. Among the viable alternatives, the TM1 mixing pattern $U^{}_{\rm TM1}$, which preserves the first column of $U^{}_{\rm TBM}$ while modifying the other two columns, provides a simple and predictive framework consistent with current experimental results~\cite{global1,global2}. Moreover, the TM1 mixing pattern has received further support from precision measurements by JUNO~\cite{JUNO:2025gmd} and has attracted renewed interest recently~\cite{Xing:2025bdm}--\cite{Ardakanian:2026rwz}.

For the form of $M^{}_{\rm D}$ in Eq.~(\ref{1.3}), in order to realize the TM1 mixing pattern, $M^{}_{\rm R}$ should take the following generic form:
\begin{eqnarray}
M^{}_{\rm R} = \left( \begin{array}{ccc} a & 2 \,b & 2 \,c \\
2 \,b &  4 \,b + d & a - b - c - d \\
2 \,c & a - b - c - d & 4 \,c + d \\
      \end{array}\right)  \;.
\label{2.2}
\end{eqnarray}
With the help of Eq.~(\ref{1.5}), it is straightforward to verify that such a form of $M^{}_{\rm R}$, in combination with an $M^{}_{\rm D}$ of the form in Eq.~(\ref{1.3}), indeed yields an $M^{}_\nu$ that is also of the form in Eq.~(\ref{2.2}) and thus leads to the TM1 mixing pattern. An illustrative realization of these mass structures within the LRSM, based on an $S_4 \times Z_5$ flavor symmetry, is presented in Appendix~A.
In the meantime, by utilizing the fact that $M^{}_\nu$ can be partially reconstructed from the experimentally measured neutrino mixing angles and mass-squared differences, the parameters $a$, $b$, $c$, and $d$ of $M^{}_{\rm R}$ can be determined numerically. Then, the obtained $M^{}_{\rm R}$ matrix allows us to proceed with the subsequent leptogenesis calculations. To obtain the right-handed neutrino masses and facilitate the leptogenesis calculations, we transform the RHN mass matrix into a diagonal form via the unitary transformation $V$:
\begin{eqnarray}
V^T M^{}_{\rm R} V =  {\rm diag}(M^{}_1, M^{}_2, M^{}_3) \;,
\label{2.3}
\end{eqnarray}
where $M^{}_I$ are the three RHN masses. In the meantime, the Dirac neutrino mass matrix is transformed to the following form:
\begin{eqnarray}
M^{\prime}_{\rm D} = M^{}_{\rm D} V \;.
\label{2.4}
\end{eqnarray}

\begin{figure*}
\centering
\includegraphics[width=5.2in]{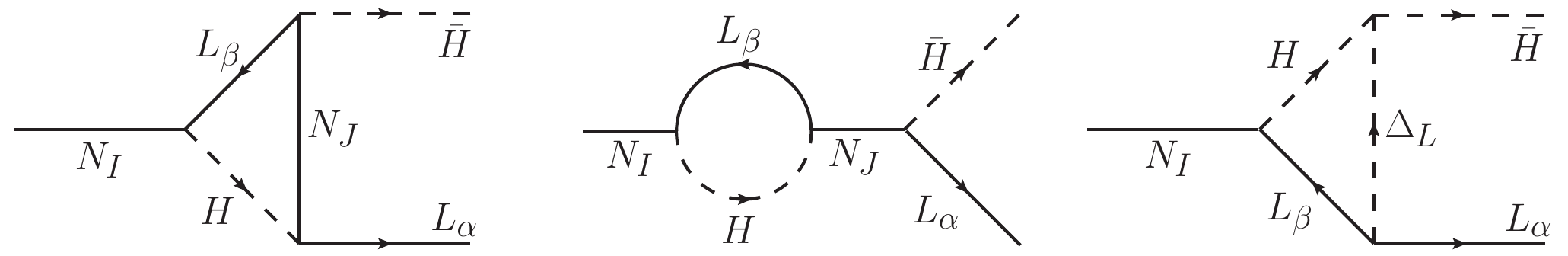}
\caption{ One-loop diagrams contributing to the CP asymmetry of $N^{}_I$ decays.  }
\label{fig1}
\end{figure*}

In the conventional leptogenesis scenario, the amount of lepton-antilepton asymmetry produced depends crucially on the CP asymmetries between the RHN decay processes $N^{}_I \to L^{}_\alpha H$ (for $\alpha = e, \mu, \tau$) and their CP-conjugate processes $N^{}_I \to \overline{L}^{}_\alpha \overline{H}$. In the case that the RHN masses are hierarchical, the flavor-specific CP asymmetries of $N^{}_I$ decays are given by
\begin{eqnarray}
&& \varepsilon^{N}_{I \alpha} = \frac{1}{8\pi (M^{\prime \dagger}_{\rm D}
M^{\prime}_{\rm D})^{}_{II} v^2} \sum^{}_{J \neq I} \left\{ {\rm Im}\left[(M^{\prime *}_{\rm D})^{}_{\alpha I} (M^{\prime}_{\rm D})^{}_{\alpha J}
(M^{\prime \dagger}_{\rm D} M^{\prime}_{\rm D})^{}_{IJ}\right] {\cal F} \left( \frac{M^2_J}{M^2_I} \right) \right. \nonumber \\
&& \hspace{1.cm}
+ \left. {\rm Im}\left[(M^{\prime *}_{\rm D})^{}_{\alpha I} (M^{\prime}_{\rm D})^{}_{\alpha J} (M^{\prime \dagger}_{\rm D} M^{\prime }_{\rm D})^*_{IJ}\right] {\cal G}  \left( \frac{M^2_J}{M^2_I} \right) \right\} \; ,
\label{2.5}
\end{eqnarray}
with ${\cal F}(x) = \sqrt{x} \left\{ (2-x)/(1-x) + (1+x) \ln [x/(1+x)] \right\}$ and ${\cal G}(x) = 1/(1-x)$. This contribution originates from the interference between the tree-level and the one-loop diagrams mediated by RHNs (see the first two diagrams in Figure~1). For the form of $M^{}_{\rm D}$ in Eq.~(\ref{1.3}), $M^{\prime}_{\rm D}$ in Eq.~(\ref{2.4}) is simply proportional to (up to a column permutation) the unitary matrix $V$ that diagonalizes $M^{}_{\rm R}$. This relation enforces orthogonality among the different columns of $M^{\prime}_{\rm D}$ (i.e., $(M^{\prime \dagger}_{\rm D} M^{\prime}_{\rm D})^{}_{IJ}=0$), which consequently makes $\varepsilon^{N}_{I \alpha}$ in Eq.~(\ref{2.5}) vanish. Therefore, the conventional leptogenesis mechanism is forbidden from working.

In the LRSM, thanks to the interactions involving the Higgs triplet $\Delta^{}_{\rm L}$, the CP asymmetry of $N^{}_I$ decays receives an additional contribution mediated by $\Delta^{}_{\rm L}$ (see the third diagram in Figure~\ref{fig1}) given by~\cite{Hambye:2003ka, Antusch:2004xy}
\begin{eqnarray}
\varepsilon^\Delta_{I \alpha}=-\frac{1}{2 \pi} \frac{\sum_\beta
{\rm Im}\left[(M^{\prime *}_{\rm D})^{}_{\alpha I}
(M^{\prime *}_{\rm D})^{}_{\beta I} f^*_{\alpha \beta} \mu^*\right]}{(M^{\prime \dagger}_{\rm D}
M^{\prime}_{\rm D})^{}_{II} M^{}_I} \,
\left[1-\frac{M^2_\Delta}{M^2_{I}} \ln \left(1+\frac{M^2_I}{M^2_{\Delta}} \right) \right] \,.
\label{2.6}
\end{eqnarray}
Here, $M_\Delta$ denotes the $\Delta^{}_{\rm L}$ mass, and $\mu$ is the dimensional coupling parameter of the trilinear term $\mu H^T i\sigma_2 \Delta^\dagger_{\rm L} H$ with $\sigma^{}_2$ being the second Pauli matrix. It can be related to the $\Delta^{}_{\rm L}$ VEV $v^{}_{\rm L}$ via $\mu = v^{}_{\rm L} M^2_\Delta / v^2$~\cite{Du:2018eaw}.
Unlike the contribution in Eq.~(\ref{2.5}), this contribution depends on a different flavor structure involving the triplet Yukawa coupling matrix $f$, and is not constrained to vanish by the orthogonality condition $(M^{\prime\dagger}_{\rm D}M^\prime_{\rm D})_{IJ}=0$.
Therefore, it holds the key to rescuing leptogenesis. Combining both contributions, the total CP asymmetry of $N^{}_I$ decays is given by
\begin{eqnarray}
\varepsilon^{}_{I \alpha} = \varepsilon^{N}_{I \alpha} + \varepsilon^\Delta_{I \alpha} \;.
\label{2.7}
\end{eqnarray}
However, it should be kept in mind that for the form of $M^{}_{\rm D}$ (and $M^{}_{\rm R}$) in Eq.~(\ref{1.3}), $\varepsilon^{N}_{I \alpha}$ always vanishes.

To be accurate, we will include flavor effects in our leptogenesis calculations. As is known, depending on the temperature range in which leptogenesis takes place (approximately the RHN mass scale), the following three distinct leptogenesis regimes can be identified~\cite{flavor1, flavor2}.
(1) Unflavored regime: In the temperature range above $10^{12}$ GeV, where the charged-lepton Yukawa $y^{}_\alpha$ interactions are not in thermal equilibrium, the lepton flavors are indistinguishable. Consequently, the three lepton flavors should be treated in a universal way. In the case of hierarchical RHN masses, the final baryon asymmetry mainly comes from the lightest RHN $N^{}_1$, since its related processes effectively wash out the lepton asymmetries generated from the heavier RHNs. The final baryon asymmetry from $N^{}_1$ can be calculated according to
\begin{eqnarray}
Y^{}_{\rm B} = c Y^{}_{\rm L} = c d \varepsilon^{}_1 \kappa(\widetilde m^{}_1)  \;,
\label{2.8}
\end{eqnarray}
where $c \simeq -1/3$ is the conversion efficiency from the lepton asymmetry to the baryon asymmetry via sphaleron processes, and $d \simeq 4 \times 10^{-3}$ measures the ratio of the equilibrium number density of $N^{}_1$ to the entropy density. The total CP asymmetry $\varepsilon^{}_1$ is obtained by summing the contributions from the three lepton flavors (i.e., $\varepsilon^{}_I = \sum^{}_\alpha \varepsilon^{}_{I \alpha}$). Finally, $\kappa(\widetilde m^{}_1)$ is the efficiency factor (i.e., the survival probability of the lepton asymmetry generated from $N^{}_1$ decays), which takes account of the washout effects due to the inverse decays of $N^{}_1$ and various lepton-number-violating scattering processes. Its concrete value depends on the washout mass parameter
\begin{eqnarray}
\widetilde m^{}_I = \sum^{}_\alpha \widetilde m^{}_{I \alpha} = \sum^{}_\alpha  \frac{|(M^{\prime}_{\rm D})^{}_{\alpha I}|^2}{M^{}_I} \;,
\label{2.9}
\end{eqnarray}
and can be numerically calculated by solving the relevant Boltzmann equations.
(2) Two-flavor regime: In the temperature range $10^{9}$--$10^{12}$ GeV, where the $y^{}_\tau$-related interactions are in thermal equilibrium while those of $e$ and $\mu$ are not, the $\tau$ flavor becomes distinguishable from the other two. Therefore, there effectively exist two distinct flavors: the $\tau$ flavor and a coherent superposition of the $e$ and $\mu$ flavors. In this regime, the final baryon asymmetry from $N^{}_1$ can be calculated according to
\begin{eqnarray}
Y^{}_{\rm B} = c(Y^{}_{{\rm L}\gamma} + Y^{}_{{\rm L}\tau} )
=  c d \left[ \varepsilon^{}_{1 \gamma} \kappa \left( \widetilde m^{}_{1 \gamma} \right) + \varepsilon^{}_{1 \tau} \kappa \left(\widetilde m^{}_{1 \tau} \right) \right]
 \;,
\label{2.10}
\end{eqnarray}
with $\varepsilon^{}_{I \gamma} = \varepsilon^{}_{I e} + \varepsilon^{}_{I \mu}$ and $\widetilde m^{}_{I \gamma} = \widetilde m^{}_{I e} + \widetilde m^{}_{I \mu}$. Here $Y^{}_{{\rm L}\alpha}$ denotes the lepton asymmetry stored in the $\alpha$ flavor.
(3) Three-flavor regime: In the temperature range below $10^{9}$ GeV, where the $y^{}_\mu$-related interactions are also in thermal equilibrium, all flavors are distinguishable, and thus each flavor should be treated separately. In this regime, the final baryon asymmetry from $N^{}_1$ can be calculated according to
\begin{eqnarray}
Y^{}_{\rm B} = c(Y^{}_{{\rm L}e} + Y^{}_{{\rm L}\mu} + Y^{}_{{\rm L}\tau} ) = c d \left[ \varepsilon^{}_{1 e} \kappa \left( \widetilde m^{}_{1 e} \right) + \varepsilon^{}_{1 \mu} \kappa \left( \widetilde m^{}_{1 \mu} \right) + \varepsilon^{}_{1 \tau} \kappa \left( \widetilde m^{}_{1\tau} \right) \right] \; .
\label{2.11}
\end{eqnarray}

\subsection{Realization of leptogenesis}

\begin{figure*}
\centering
\includegraphics[width=6.5in]{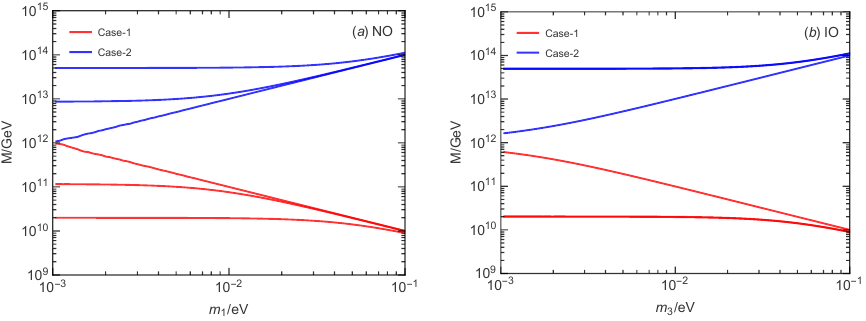}
\caption{ In the NO (IO) case, for the benchmark value of $r$, the possible values of three right-handed neutrino masses as functions of the lightest neutrino mass $m^{}_1$ ($m^{}_3$) for a representative parameter combination (see the main text for details).}
\label{fig2}
\end{figure*}

For $M^{}_{\rm D}$ in Eq.~(\ref{1.3}) and $M^{}_{\rm R}$ in Eq.~(\ref{2.2}), both the type-I and type-II seesaw contributions to $M^{}_\nu$ individually exhibit the TM1 form in Eq.~(\ref{2.2}). As a result, the reconstruction of $M^{}_{\rm R}$ from $M^{}_\nu$ (which contains the low-energy neutrino observables) via Eq.~(\ref{1.5}) admits two distinct scenarios, related by an interchange of the type-I and type-II seesaw contributions while keeping the total $M^{}_\nu$ unchanged. We refer to these scenarios as Case-1 and Case-2. In Case-1, the type-II seesaw contribution becomes progressively less important as $r$ decreases, whereas in Case-2 it is the type-I seesaw contribution that is gradually suppressed for smaller values of $r$. Both cases reproduce the same low-energy neutrino observables but can lead to different RHN mass spectrum and leptogenesis predictions.

Since the concrete realization of leptogenesis highly depends on the RHN mass spectrum, it is useful to first examine the latter obtained from the reconstruction. Using the experimental measurements of the neutrino mixing angles and mass-squared differences as inputs, we show in Figure~\ref{fig2} (a) and (b) (for the normal ordering (NO) and inverted ordering (IO) cases of light neutrino masses, respectively) the possible values of the three RHN masses as functions of the lightest neutrino mass ($m^{}_1$ and $m^{}_3$, respectively). In obtaining these results, the two Majorana CP phases of the neutrino mixing matrix are fixed at $\rho=\sigma=0$ ($\rho$ and $\sigma$ are associated with $m_1$ and $m_2$, respectively), while the Dirac CP phase is constrained by the TM1 mixing scheme considered in the present paper. We find that varying the Majorana CP phases does not alter the qualitative features of the RHN mass spectrum shown in Figure~\ref{fig2}.
For the purpose of illustration, we have taken $r=10^{-24}$ and $m^{}_{\rm D} = 1$ GeV as typical inputs. For other choices of $m^{}_{\rm D}$ and $r$, the RHN masses adjust accordingly to reproduce the low-energy neutrino observables. For example, increasing $m^{}_{\rm D}$ by one order of magnitude while decreasing $r$ by two orders of magnitude raises the RHN masses by two orders of magnitude. This can be easily understood with the help of Eq.~(\ref{1.5}).

As shown in Figure~\ref{fig2}, the RHN mass spectrum exhibits two qualitatively different patterns. In the first scenario, the spectrum is strongly hierarchical, $M^{}_1 \ll M^{}_2, M^{}_3$, so that the baryon asymmetry is dominantly generated by the lightest RHN $N^{}_1$. Its contribution can be calculated using the formalism in Eqs.~(\ref{2.8})--(\ref{2.11}), with the appropriate flavor regime taken into account. In the second scenario, the spectrum satisfies $M^{}_1 \approx M^{}_2 \ll M^{}_3$, such that both $N^{}_1$ and $N^{}_2$ contribute significantly to the final baryon asymmetry. In this case, the total baryon asymmetry receives comparable contributions from both $N^{}_1$ and $N^{}_2$, and is obtained by including the washout effects associated with both $N^{}_1$ and $N^{}_2$ simultaneously. The resulting baryon asymmetry can be expressed as
\begin{eqnarray}
Y^{}_{\rm B}  = c d (\varepsilon^{}_{1} + \varepsilon^{}_{2}) \kappa \left( \widetilde m^{}_1 +  \widetilde m^{}_2 \right) \;,
\label{2.12}
\end{eqnarray}
in the unflavored regime, or
\begin{eqnarray}
Y^{}_{\rm B}  = c d \left[ (\varepsilon^{}_{1 \gamma} + \varepsilon^{}_{2 \gamma}) \kappa \left( \widetilde m^{}_{1 \gamma} +  \widetilde m^{}_{2 \gamma} \right) + (\varepsilon^{}_{1 \tau} + \varepsilon^{}_{2 \tau}) \kappa \left( \widetilde m^{}_{1 \tau} +  \widetilde m^{}_{2 \tau} \right) \right]\;,
\label{2.13}
\end{eqnarray}
in the two-flavor regime, or
\begin{eqnarray}
Y^{}_{\rm B}  = c d \left[ (\varepsilon^{}_{1 e} + \varepsilon^{}_{2 e}) \kappa \left( \widetilde m^{}_{1 e} +  \widetilde m^{}_{2 e} \right) + (\varepsilon^{}_{1 \mu} + \varepsilon^{}_{2 \mu}) \kappa \left( \widetilde m^{}_{1 \mu} +  \widetilde m^{}_{2 \mu} \right)  + (\varepsilon^{}_{1 \tau} + \varepsilon^{}_{2 \tau}) \kappa \left( \widetilde m^{}_{1 \tau} +  \widetilde m^{}_{2 \tau} \right) \right]\;,
\label{2.14}
\end{eqnarray}
in the three-flavor regime.

\begin{figure*}
\centering
\includegraphics[width=6.5in]{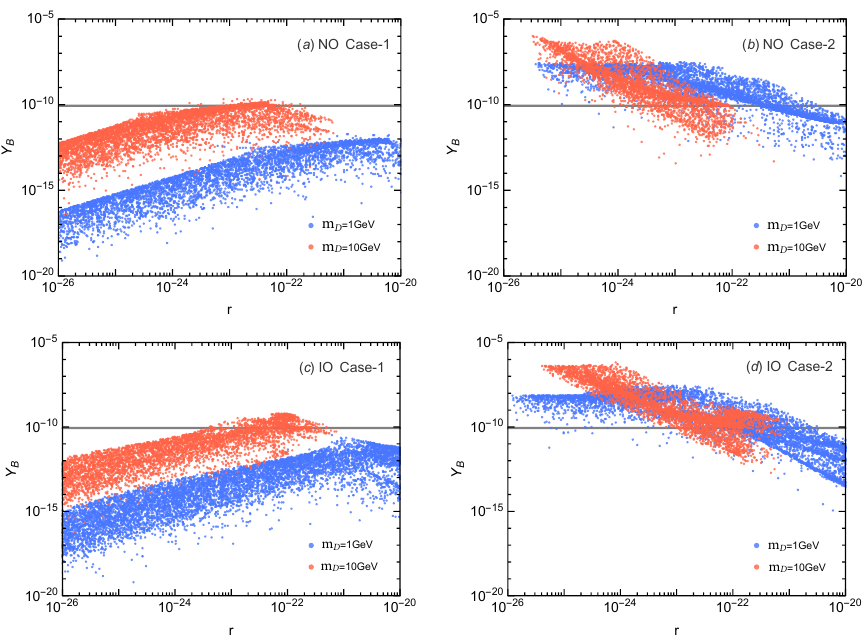}
\caption{ For the scenario studied in section~2, in the NO Case-1, NO Case-2, IO Case-1, and IO Case-2 scenarios, the allowed values of $Y^{}_{\rm B}$ as functions of $r$ for some benchmark values of $m^{}_{\rm D}$. The horizontal line stands for the observed value of $Y^{}_{\rm B}$.}
\label{fig3}
\end{figure*}

Now, we are ready to perform the numerical calculations of leptogenesis.
In Figures~\ref{fig3}(a)--(d) (for the NO Case-1, NO Case-2, IO Case-1, and IO Case-2 scenarios, respectively), we show the allowed values of $Y^{}_{\rm B}$ as functions of $r$ for the benchmark values $m^{}_{\rm D} = 1$ and $10$ GeV. These results are obtained with the following parameter settings: for the neutrino mixing angles and mass-squared differences, we employ the global-fit results from Refs.~\cite{global1, global2}. The two Majorana CP phases of the neutrino mixing matrix are allowed to vary in the range $0$--$2\pi$, while the Dirac CP phase is constrained by the TM1 mixing scheme considered in the present paper. The lightest neutrino mass ($m^{}_1$ or $m^{}_3$ in the NO or IO case) is varied between $0.001$ eV and $0.1$ eV. We have taken $M^{}_\Delta = 10 M^{}_1$ as a benchmark value. In fact, the final result is approximately independent of the concrete ratio $M^{}_\Delta/M^{}_I$ provided that $\Delta^{}_{\rm L}$ is much heavier than the RHNs. This can be easily understood from the following fact: for a small $M^{}_I/M^{}_\Delta$, $\varepsilon^\Delta_{I \alpha}$ in Eq.~(\ref{2.6}) approximates to
\begin{eqnarray}
\varepsilon^\Delta_{I \alpha} \simeq -\frac{1}{8 \pi} \frac{\sum_\beta
{\rm Im}\left[(M^{\prime *}_{\rm D})^{}_{\alpha I}
(M^{\prime *}_{\rm D})^{}_{\beta I} (M^{\rm II *}_{\nu})^{}_{\alpha \beta} \right]}{(M^{\prime \dagger}_{\rm D}
M^{\prime}_{\rm D})^{}_{II} v^2 } M^{}_I \;,
\label{2.15}
\end{eqnarray}
which is independent of $M^{}_\Delta$.
Moreover, $r$ is severely constrained by light neutrino mass generation, and parameter regions requiring fine-tuned cancellations between sizable type-I and type-II seesaw contributions to yield a much smaller $M^{}_\nu$ are excluded
\footnote{To quantify the magnitude of light neutrino masses, we introduce the dimensional parameter $\overline{m} \equiv \sqrt{{\rm Tr} \left( m^{\dagger} m \right)}$. In our numerical analysis, we require both the type-I and type-II seesaw contributions to satisfy $\overline{M^{I}_\nu} < 3\,\overline{M^{}_\nu}$ and $\overline{M^{II}_\nu} < 3\,\overline{M^{}_\nu}$, thereby avoiding parameter regions in which the observed light neutrino masses arise from strong cancellations between two much larger contributions.}.
In addition, we restrict the lightest RHN mass to $M^{}_1 \leq 10^{14}$ GeV, since for larger masses non-resonant $\Delta L=2$ scattering processes enter thermal equilibrium and induce a strong suppression of the efficiency factor~\cite{Giudice:2003jh}.

The results show that, for certain values of $r$, the observed value of $Y^{}_{\rm B}$ can be successfully reproduced in both the NO and IO cases. For Case-1, shown in Figures~\ref{fig3}(a) and (c) for the NO and IO cases respectively, the allowed values of $Y^{}_{\rm B}$ decrease as $r$ becomes smaller. This trend arises because a smaller value of $r$ suppresses the type-II seesaw contribution to the light neutrino mass matrix, thereby reducing the $\Delta^{}_{\rm L}$-mediated CP asymmetry. In contrast, the results for Case-2, shown in Figures~\ref{fig3}(b) and (d), exhibit the opposite behavior: the allowed values of $Y^{}_{\rm B}$ increase as $r$ decreases. This difference originates from the above-mentioned interchange of the type-I and type-II seesaw contributions in the reconstruction of $M^{}_{\rm R}$ from $M^{}_\nu$. As can be seen from Eq.~(\ref{2.15}), the $\Delta^{}_{\rm L}$-induced CP asymmetry is directly proportional to the type-II seesaw contribution. Consequently, the opposite evolution of the type-II component with $r$ in the two cases naturally results in the opposite behavior of the baryon asymmetry.
Figure~\ref{fig3} also demonstrates that the generated baryon asymmetry depends sensitively on the overall scale $m^{}_{\rm D}$ of the Dirac neutrino mass matrix, for both Case-1 and Case-2. This is because larger values of $m^{}_{\rm D}$ correspond to larger RHN masses, as discussed above, which are subsequently helpful for enhancing the CP asymmetry of RHN decays as shown in Eq.~(\ref{2.15}).
Importantly, successful leptogenesis can still be achieved even when the type-II seesaw contribution to $M^{}_\nu$ is subdominant. These results demonstrate that the $\Delta^{}_{\rm L}$-mediated contribution to the CP asymmetry of RHN decays can by itself rescue leptogenesis that is forbidden in the conventional type-I seesaw framework.

\begin{figure*}
\centering
\includegraphics[width=6.5in]{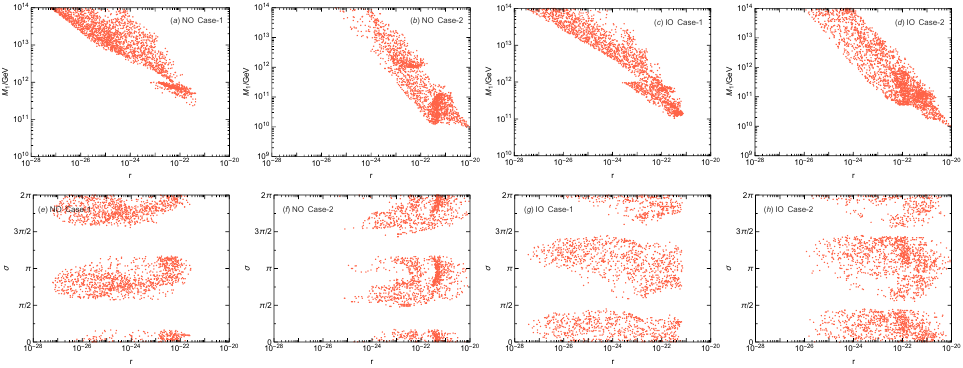}
\caption{ For the scenario studied in section~2, in the NO Case-1, NO Case-2, IO Case-1, and IO Case-2 scenarios, the values of the lightest RHN mass and the Majorana CP phase $\sigma$ that allow for a reproduction of the observed value of $Y^{}_{\rm B}$ as functions of $r$.}
\label{fig4}
\end{figure*}

For these scenarios, in Figures~\ref{fig4}(a)--(d) we further show the values of $M^{}_1$ (as functions of the ratio $r$) that allow for successful leptogenesis. With the same parameter settings as above, these results are obtained with $m^{}_{\rm D}$ varying in the range from 0.1 to 100 GeV. The results show that the corresponding lower bound on $M^{}_1$ is of order $10^{10}$ GeV for successful leptogenesis to be achieved. Furthermore, Figures~\ref{fig4}(e)--(h) show the parameter space of $\sigma$ versus the ratio $r$ for successful leptogenesis (the Majorana phase $\rho$ does not exhibit significant constraints and is therefore not shown).
One can see that the allowed values of $\sigma$ are distributed in two disconnected bands. In the NO case, these bands are approximately in $\pi/2$ to $5\pi/4$ and $3\pi/2$ to $2\pi$, while in the IO case the allowed region is somewhat broader.

\section{Leptogenesis for the exactly degenerate form of $M_{\rm R}$}

In this section, we investigate the realization of leptogenesis within the LRSM for the exactly degenerate form of $M^{}_{\rm R}$ in Eq.~(\ref{1.3}), while $M^{}_{\rm D}$ acquires a nontrivial flavor structure from certain symmetry-breaking patterns of the employed flavor symmetries. We also take the resulting neutrino mixing to be of the TM1 type as an illustration. Specifically, $M^{}_{\rm D}$ takes the same form as in Eq.~(\ref{2.2}).

With the help of Eq.~(\ref{1.5}), it is straightforward to verify that an $M^{}_{\rm D}$ of the form in Eq.~(\ref{2.2}), in combination with an $M^{}_{\rm R}$ of the form in Eq.~(\ref{1.3}), indeed yields an $M^{}_\nu$ that is also of the form in Eq.~(\ref{2.2}) and thus leads to the TM1 mixing pattern.
In the meantime, by utilizing the fact that $M^{}_\nu$ can be partially reconstructed from the experimentally measured neutrino mixing angles and mass-squared differences, the parameters $a$, $b$, $c$, and $d$ of $M^{}_{\rm D}$ can be determined numerically. Then, the obtained $M^{}_{\rm D}$ matrix allows us to proceed with the subsequent leptogenesis calculations. Unlike the scenario discussed in the previous section, the reconstruction of $M^{}_{\rm D}$ from $M^{}_\nu$ in the present scenario does not admit two distinct cases related by an interchange of the type-I and type-II seesaw contributions while keeping the total $M^{}_\nu$ unchanged.
For this reason, only Case-1 (defined at the beginning of Section~2.2) can be realized in the present scenario.

In the case that the RHN masses are nearly degenerate, there are two important differences for the leptogenesis calculations. First, the RHN-mediated CP asymmetries of $N^{}_I$ decays become resonantly enhanced as~\cite{resonant1, resonant2}
\begin{eqnarray}
\varepsilon^{N}_{I\alpha} = \frac{{\rm Im}\left\{ (M^{\prime *}_{\rm D})^{}_{\alpha I} (M^{\prime}_{\rm D})^{}_{\alpha J}
\left[ M^{}_J (M^{\prime \dagger}_{\rm D} M^{\prime}_{\rm D})^{}_{IJ} + M^{}_I (M^{\prime \dagger}_{\rm D} M^{\prime}_{\rm D})^{}_{JI} \right] \right\} }{8\pi  v^2 (M^{\prime \dagger}_{\rm D} M^{\prime}_{\rm D})^{}_{II}} \cdot \frac{M^{}_I \Delta M^2_{IJ}}{(\Delta M^2_{IJ})^2 + M^2_I \Gamma^2_J} \;,
\label{3.1}
\end{eqnarray}
where $\Delta M^2_{IJ} \equiv M^2_I - M^2_J$ and $\Gamma^{}_J = (M^{\prime \dagger}_{\rm D} M^{\prime}_{\rm D})^{}_{JJ} M^{}_J/(8\pi v^2)$ is the decay rate of $N^{}_J$ (for $J \neq I$).
Second, the contributions of all the nearly degenerate RHNs to the final baryon asymmetry are on the same footing and should be taken into consideration altogether. Therefore, for the present scenario, all three RHNs $N^{}_1$, $N^{}_2$ and $N^{}_3$ would in principle contribute significantly to the baryon asymmetry (as given by Eqs.~(\ref{2.12})--(\ref{2.14}), with the contribution of $N^{}_3$ included analogously).
However, for the form of $M^{}_{\rm R}$ in Eq.~(\ref{1.3}), the three RHNs are exactly degenerate in mass (i.e., $\Delta M_{IJ} = 0$), which makes $\varepsilon^{N}_{I\alpha}$ in Eq.~(\ref{3.1}) vanish. Therefore, the conventional leptogenesis mechanism is forbidden from working. For the form of $M_{\rm R}$ in Eq.~(\ref{1.3}), the RHN masses are exactly degenerate as a direct consequence of the flavor-symmetric structure. This is different from resonant leptogenesis, where a small but non-zero mass splitting among the RHNs is required to enhance the conventional CP asymmetry. In some flavor-symmetry models, such a splitting is attributed to higher-order contributions to the RHN mass matrix~\cite{Borah:2017qdu}. However, the small RHN mass splitting required for resonant leptogenesis demands that these higher-order contributions be strongly suppressed relative to the leading-order terms, which is difficult to justify without an additional suppression mechanism.
In the present work, we retain the exactly degenerate structure of $M_{\rm R}$, so that the conventional RHN-mediated CP asymmetry vanishes.

\begin{figure*}
\centering
\includegraphics[width=6.5in]{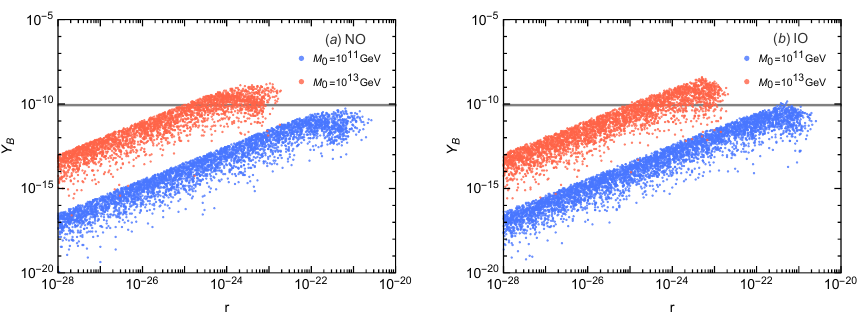}
\caption{ For the scenario studied in section~3, in the NO (IO) case, the allowed values of $Y^{}_{\rm B}$ as functions of $r$ for some benchmark values of $M^{}_0$. The horizontal line stands for the observed value of $Y^{}_{\rm B}$.}
\label{fig5}
\end{figure*}

As mentioned above, the interactions involving the Higgs triplets provide an additional contribution to the total CP asymmetry, as shown in Eq.~(\ref{2.6}). We therefore investigate whether this contribution can rescue leptogenesis.
In Figure~\ref{fig5}(a) and (b) (for the NO and IO cases, respectively), we show the allowed values of $Y^{}_{\rm B}$ as functions of $r$ for the benchmark values $M^{}_0 = 10^{11}$ and $10^{13}$ GeV. These results are obtained with the same parameter settings as those used in Figure~3 (see the paragraph following Eq.~(\ref{2.14})). The results show that in both the NO and IO cases, the observed value of $Y^{}_{\rm B}$ can be successfully reproduced. It is clear that the allowed values of $Y^{}_{\rm B}$ increase with $r$, which arises from the gradual suppression of the type-II seesaw contribution for smaller values of $r$ (see Eqs.~(\ref{1.5}) and (\ref{2.15})). In addition, a larger $M_0$ generally leads to higher values of $Y_{\rm B}$. Overall, when $M_0$ varies from $10^{11}$ to $10^{13}$ GeV, the observed baryon asymmetry can be reproduced over a broad range of $r$, approximately from $10^{-25}$ to $10^{-21}$. We further note that even in regions where the type-II seesaw contribution to $M^{}_\nu$ is relatively suppressed, the $\Delta^{}_{\rm L}$-mediated CP asymmetry remains sufficient to account for the observed baryon asymmetry, thereby rescuing leptogenesis that would otherwise be forbidden in the conventional type-I seesaw framework.

\begin{figure*}
\centering
\includegraphics[width=6.5in]{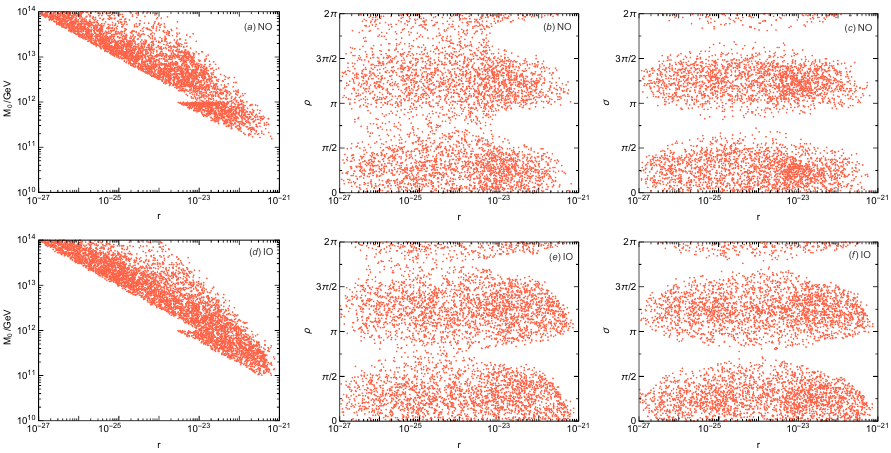}
\caption{ For the scenario studied in section~3, in the NO (IO) case, the values of the RHN mass and the two Majorana CP phases $\rho$ and $\sigma$ that allow for a reproduction of the observed value of $Y^{}_{\rm B}$ as functions of $r$.}
\label{fig6}
\end{figure*}

In Figures~\ref{fig6}(a) and (d), we further show the values of $M^{}_{0}$ (as functions of $r$) that allow for successful leptogenesis. With the same parameter settings as above, these results are obtained with $M_0$ varying in the range from $10^{11}$ to $10^{14}$ GeV. The results show that successful leptogenesis can be achieved only for $M^{}_{0} \gtrsim 10^{11}$ GeV. In our numerical analysis, $v_{\rm L}$ and $v_{\rm R}$ enter only through their ratio $r=v_{\rm L}/v_{\rm R}$ and are therefore not specified separately. The left-handed triplet VEV $v_{\rm L}$ is constrained to be at most of order GeV by electroweak precision measurements \cite{Kriewald:2024cgr}. In this work, $v_{\rm R}$ is taken to be sufficiently large such that the right-handed charged gauge boson is much heavier than the RHNs relevant for leptogenesis. This choice is consistent with the constraint on $v_{\rm L}$, given the small values of $r$ considered in our analysis. Consequently, the decay and scattering processes mediated by this heavy gauge boson are strongly suppressed and are neglected in our numerical analysis. 
Furthermore, Figures~\ref{fig6}(b), (c), (e), and (f) show the allowed parameter spaces of $\rho$ and $\sigma$ versus $r$ for successful leptogenesis. One can see that the allowed values of $\rho$ depend on $r$. For smaller values of $r$ (corresponding to larger values of $M_0$), successful leptogenesis can be achieved for almost the entire range $0$--$2\pi$. As $r$ increases, however, the allowed region gradually shrinks and becomes concentrated in two separated bands approximately located around $0$--$\pi/2$ and $\pi$--$3\pi/2$. In contrast, $\sigma$ remains significantly constrained throughout the parameter space. The allowed points are mainly distributed in two disconnected regions, roughly corresponding to $0$--$\pi/2$ and $\pi$--$3\pi/2$. Similar features are observed in both the NO and IO cases.

\section{Applicability to more general type-I+II seesaw models}

As demonstrated in Sections 2 and 3, successful leptogenesis can be achieved within the LRSM when either $M_{\rm D}$ or $M_{\rm R}$ takes the exactly degenerate form in Eq.~(\ref{1.3}). It is worth noting that the scalar-triplet contribution to the CP asymmetries of RHN decays is not restricted to the LRSM, but also arises in more general type-I+II seesaw models. In such models, in addition to three RHNs, the SM is further extended by an $SU(2)_{\rm L}$ Higgs triplet $\Delta$, which generates the type-II contribution $M_\nu^{\rm II}=2v_\Delta f_\Delta$~\cite{Magg:1980ut, Cheng:1980qt}. Consequently, the light-neutrino mass matrix is given by
\begin{eqnarray}
M_\nu=M_\nu^{\rm II}+M_\nu^{\rm I}
=2v_\Delta f_\Delta-M_{\rm D}M_{\rm R}^{-1}M_{\rm D}^T,
\label{4.1}
\end{eqnarray}
where $v_\Delta$ and $f_\Delta$ denote the VEV and Yukawa coupling matrix of $\Delta$, respectively. Unlike in the LRSM, $M_\nu^{\rm II}$ and $M_{\rm R}$ are not necessarily proportional to each other, since their corresponding Yukawa coupling matrices are independent.

For the scenario considered in Section 3, $M_{\rm R}$ takes the exactly degenerate form in Eq.~(\ref{1.3}), while $M_{\rm D}$ acquires a nontrivial flavor structure through the spontaneous breaking of the flavor symmetry. Such a scenario may be realized through a non-Abelian discrete flavor symmetry with appropriate invariant contractions, where the lepton doublets and RHNs are assigned to triplet representations, while the SM Higgs doublet and the scalar triplet $\Delta$ transform as singlets. An auxiliary discrete symmetry can be introduced to allow the flavon-independent Majorana RHN mass term and triplet Yukawa interaction, while forbidding the corresponding interactions involving flavon fields. In contrast, the neutrino Dirac Yukawa interactions involve flavon fields, whose suitable VEV alignments generate a nontrivial TM1 structure of $M_{\rm D}$. In this way, both $M_{\rm R}$ and $M_\nu^{\rm II}$ take the exactly degenerate form in Eq.~(\ref{1.3}), yielding the proportionality relation $M_\nu^{\rm II}=\xi M_{\rm R}$. The reconstruction of $M_{\rm D}$ from $M_\nu$ therefore retains the same form as in Section 3, with $r$ replaced by the independent parameter $\xi$.

For the scenario considered in Section 2, $M_{\rm D}$ takes the exactly degenerate form in Eq.~(\ref{1.3}), while both $M_{\rm R}$ and $M_\nu^{\rm II}$ acquire nontrivial flavor structures through the spontaneous breaking of the flavor symmetry. This scenario may also be realized within a similar flavor-symmetry framework, allowing the flavon-independent neutrino Dirac Yukawa interaction while requiring flavon insertions in the Majorana RHN mass terms and triplet Yukawa interactions. With suitable flavon VEV alignments, both $M_{\rm R}$ and
$M_\nu^{\rm II}$ can independently acquire nontrivial TM1 structures. Since the corresponding Yukawa couplings are independent, $M_\nu^{\rm II}$ need not be directly proportional to $M_{\rm R}$. Nevertheless, for the exactly degenerate form of $M_{\rm D}$ in Eq. (3), both the type-I and type-II seesaw contributions can still preserve the TM1 structure.

In both scenarios, the scalar triplet $\Delta$ can generate an additional contribution to the CP asymmetries of RHN decays, analogous to the $\Delta_{\rm L}$-mediated contribution in the LRSM. Although the conventional RHN-mediated CP asymmetries vanish due to the exactly degenerate form of either $M_{\rm D}$ or $M_{\rm R}$, the scalar-triplet contribution can rescue the leptogenesis as in the LRSM.

\section{Summary}

While the type-I seesaw model provides a unified framework for explaining the origin of neutrino masses and the baryon asymmetry of the Universe, flavor symmetries offer an attractive approach to understanding the observed neutrino mixing pattern. In many type-I seesaw models based on non-Abelian flavor symmetries, the three lepton doublets and the three right-handed neutrinos are organized into a triplet representation, respectively. In the symmetry limit, this assignment naturally leads to highly constrained neutrino mass matrices as shown in Eq.~(\ref{1.3}).
Unfortunately, such forms of $M^{}_{\rm D}$ and $M^{}_{\rm R}$ forbid the conventional leptogenesis mechanism from working, either because of the orthogonality relations among different columns of the neutrino Yukawa coupling matrix or due to the exact degeneracy of three RHNs.

In this paper, we have explored alternative mechanisms that can rescue leptogenesis for the forms of $M^{}_{\rm D}$ and $M^{}_{\rm R}$ in Eq.~(\ref{1.3}), without breaking the original flavor structure dictated by the employed flavor symmetries.
This provides a complementary possibility to scenarios in which additional flavor symmetry breaking is introduced to render the conventional RHN-mediated CP asymmetry non-vanishing. We find that the LRSM provides a particularly appealing framework for this purpose. Specifically, the Higgs triplet $\Delta^{}_{\rm L}$ present in this model induces an additional contribution to the CP asymmetry of RHN decays. We therefore investigate whether this contribution can rescue leptogenesis. To be specific, we have considered two scenarios: either $M^{}_{\rm D}$ or $M^{}_{\rm R}$ takes the form in Eq.~(\ref{1.3}), while the other acquires a nontrivial flavor structure from certain breaking patterns of the employed flavor symmetries. For definiteness, we have taken the resulting neutrino mixing pattern to be of the TM1 type as an illustration.

For an $M^{}_{\rm D}$ of the form in Eq.~(\ref{1.3}), two distinct scenarios for the reconstruction of $M^{}_{\rm R}$ from $M^{}_\nu$ have been identified, corresponding to different realizations of the type-I and type-II seesaw contributions. Although both scenarios reproduce the same low-energy neutrino observables, they lead to qualitatively different leptogenesis predictions. In particular, the generated baryon asymmetry exhibits opposite dependencies on $r$. The observed value of $Y_{\rm B}$ can be successfully reproduced for both the NO and IO cases, with the viable parameter space implying a lower bound on the lightest RHN mass at approximately the $10^{10}$ GeV scale. The requirement of successful leptogenesis also imposes constraints on the Majorana CP phase $\sigma$.

For an $M^{}_{\rm R}$ of the form in Eq.~(\ref{1.3}), successful leptogenesis can be achieved over a broad range of $r$, with the corresponding RHN mass scale required to be above approximately $10^{11}$ GeV. The viable parameter space further imposes constraints on the Majorana CP phases. In particular, the phase $\sigma$ remains significantly restricted, while the allowed range of $\rho$ becomes progressively broader as $r$ decreases.

In summary, our results demonstrate that the Higgs triplet of the LRSM provides a natural and viable additional CP-violating contribution that can rescue conventional leptogenesis in the exactly degenerate scenarios considered here. While such a scalar-triplet contribution is not unique to the LRSM, the mechanism considered here may also be applied to more general type-I+II seesaw models. However, the LRSM is particularly attractive for two reasons: first, the RHNs and Higgs triplets $\Delta_{L,R}$ exist naturally in the LRSM, allowing the type-I and type-II seesaw contributions to be simultaneously realized within the same framework; second, the type-II contribution is directly related to the Majorana RHN mass matrix. In more general type-I+II seesaw models, this proportionality relation is not required, and $M_\nu^{\rm II}$ and $M_{\rm R}$ generally involve independent parameters. We also emphasize that the TM1 realization adopted in our numerical analysis serves only as a concrete illustration. More generally, the mechanism considered here may also be applicable to other textures of the neutrino mass matrices that preserve the corresponding conditions responsible for the vanishing of conventional RHN-mediated CP asymmetry.

\appendix
\section{An $S_4$ realization in the LRSM}

To illustrate the mass structures considered in Section~2, we present an $S_4$ flavor-symmetry realization within the LRSM. As pointed out in Ref.~\cite{Rodejohann:2015hka}, the left-right gauge structure imposes additional constraints on flavor assignments compared with conventional type-I seesaw models. In particular, the right-handed neutrino and charged lepton belong to the same $SU(2)_R$ doublet and therefore share the same flavor assignment. In our construction, the left- and right-handed lepton doublets $\ell_L=(\nu_L,e_L)^T$ and $\ell_R=(\nu_R,e_R)^T$ are assigned to equivalent $\mathbf{3}$ representations of $S_4$. The Higgs sector consists of one bidoublet $\Phi$ and two triplets $\Delta_{\rm L}$ and $\Delta_{\rm R}$, all transforming as singlets under $S_4$. The corresponding field assignments are summarized in Table~\ref{tab:flavor}.

The bidoublet Yukawa interactions generate $M_{\rm D}=\kappa Y+\kappa'\widetilde Y$ and $M_\ell=\kappa'Y+\kappa\widetilde Y$. Here, $\kappa$ and $\kappa'$ denote the two neutral VEVs of $\Phi$,
while $Y$ and $\widetilde{Y}$ are the Yukawa coupling matrices associated with $\Phi$ and its conjugate
$\widetilde{\Phi}=\sigma_2\Phi^*\sigma_2$, respectively. An auxiliary $Z_5$ symmetry is introduced to constrain the Yukawa interactions.
It allows the Dirac Yukawa term $y_{\rm D}[\overline{\ell}_L\Phi\ell_R]_{\mathbf{1}}$, while forbidding the corresponding flavon-independent term involving $\widetilde\Phi$. In the limit $\kappa'=0$, consistent with the commonly considered hierarchy $\kappa'\ll\kappa$, the mass relations reduce to $M_{\rm D}=\kappa Y$ and $M_\ell=\kappa\widetilde Y$. The $S_4$ singlet contraction then yields the exactly degenerate form of $M_{\rm D}$ in Eq.~(\ref{1.3}), with $m_{\rm D}=\kappa y_{\rm D}$. The charged-lepton masses instead arise from three flavon multiplets $\chi_{1,2,3}$, which are singlets under the LRSM gauge group and transform as $\mathbf{1}$, $\mathbf{3'}$, and $\mathbf{3}$ under $S_4$, respectively. Their effective Yukawa interactions are $y_i^\ell[\overline{\ell}_L\widetilde\Phi\ell_R\chi_i]_{\mathbf{1}}/\Lambda$, where $\Lambda$ denotes the flavor scale. For the VEV alignments $\langle\chi_1\rangle=u_1$, $\langle\chi_2\rangle=u_2(1,0,0)^T$, and $\langle\chi_3\rangle=u_3(1,0,0)^T$, the Clebsch--Gordan coefficients given in Ref.~\cite{King:2011zj} yield three independent charged-lepton masses, with $M_\ell M_\ell^\dagger$ diagonal.

To obtain the nontrivial TM1 structure of $M_{\rm R}$, we adopt the $S_4$ flavor structure proposed in Ref.~\cite{Luhn:2013lkn}. Four flavon fields $\phi_{1,2,3,4}$, transforming as $\mathbf{1}$, $\mathbf{2}$, $\mathbf{3'}$, and $\mathbf{3'}$ under $S_4$, respectively, enter the Yukawa interactions involving the right-handed Higgs triplet $\Delta_{\rm R}$. The corresponding effective operators are
\begin{equation}
-\mathcal{L}_{R}\supset
\sum_{k=1}^{4}\frac{\lambda_k}{\Lambda}
\left[\ell_R^T C i\tau_2\Delta_{\rm R}\ell_R\phi_k\right]_{\mathbf{1}}
+\mathrm{h.c.},
\label{A.1}
\end{equation}
where $\lambda_k$ denote the effective couplings. The corresponding left-handed interactions involve $\phi_k^\dagger$, as required by the $Z_5$ symmetry, with their Yukawa couplings related by left-right parity symmetry. For real flavon VEVs and the corresponding $S_4$ contractions, the two sectors yield $f_{\rm L}=f_{\rm R}\equiv f$. Following Ref.~\cite{Luhn:2013lkn}, the flavon VEV alignments are
\begin{equation}
\langle\phi_1\rangle=v_1,\quad
\langle\phi_2\rangle=v_2(1,1)^T,\quad
\langle\phi_3\rangle=v_3(1,1,1)^T,\quad
\langle\phi_4\rangle=v_4(0,1,-1)^T.
\label{A.2}
\end{equation}
The first three flavon contributions preserve the Klein symmetry $Z_2^S\times Z_2^U$, while the fourth breaks it to the residual $Z_2^{SU}$ symmetry, generated by the product $SU$ of the $S_4$ generators $S$ and $U$. This residual symmetry preserves the first column of the tri-bimaximal mixing matrix, leading to the TM1 mixing pattern. After $\Delta_{\rm R}$ and the flavons acquire their VEVs, the RHN mass matrix takes the form
\begin{eqnarray}
M_{\rm R}={}x_1
\begin{pmatrix}
1&0&0\\
0&0&1\\
0&1&0
\end{pmatrix}
+x_2
\begin{pmatrix}
0&1&1\\
1&1&0\\
1&0&1
\end{pmatrix}
+x_3
\begin{pmatrix}
2&-1&-1\\
-1&2&-1\\
-1&-1&2
\end{pmatrix}
+y
\begin{pmatrix}
0&1&-1\\
1&2&0\\
-1&0&-2
\end{pmatrix},
\label{A.3}
\end{eqnarray}
where $x_{1,2,3}$ and $y$ are generally complex parameters determined by the corresponding couplings and VEVs. The resulting $M_{\rm R}$ reproduces the generic TM1 form in Eq.~(\ref{2.2}), with $a=x_1+2x_3$, $b=(x_2-x_3+y)/2$, $c=(x_2-x_3-y)/2$, and $d=4x_3-x_2$. The common triplet Yukawa coupling gives $M_{\rm L}=2v_{\rm L}f$ and $M_{\rm R}=2v_{\rm R}f$, yielding $M_{\rm L}=(v_{\rm L}/v_{\rm R})M_{\rm R}$, as used in Eqs.~(\ref{1.4}) and (\ref{1.5}). The required nonzero $v_{\rm L}$ can be induced by the $S_4\times Z_5$ invariant scalar interaction
\begin{equation}
V\supset
-\frac{\beta_\chi}{\Lambda}\chi_1
\operatorname{Tr}\left(
\widetilde\Phi\Delta_{\rm R}
\Phi^\dagger\Delta_{\rm L}^\dagger
\right)+\mathrm{h.c.},
\label{A.4}
\end{equation}
where $\beta_\chi$ denotes a dimensionless coupling constant.
After $\chi_1$ and $\Delta_{\rm R}$ acquire their VEVs, this interaction generates the trilinear coupling $\mu H^T i\sigma_2\Delta_{\rm L}^\dagger H$, with $\mu=\beta_\chi u_1v_{\rm R}/\Lambda$. Following electroweak symmetry breaking, this coupling induces a nonzero $v_{\rm L}$ through the minimization of the scalar potential, with $\mu=v_{\rm L}M_\Delta^2/v^2$. The resulting $v_{\rm L}$ generates the type-II contribution to the light-neutrino mass matrix, while the trilinear coupling $\mu$ also enters the $\Delta_{\rm L}$-mediated CP asymmetry discussed in Section~2.

\begin{table}[t]
\centering
\caption{Field assignments under the LRSM gauge group and the $S_4\times Z_5$ flavor symmetry. The $Z_5$ charges are given modulo five.}
\label{tab:flavor}
\setlength{\tabcolsep}{3pt}
\scriptsize
\begin{tabular}{c|cccccccccccc}
\hline\hline
 & $\ell_L$ & $\ell_R$ & $\Phi$ & $\Delta_L$ & $\Delta_R$ & $\phi_1$ & $\phi_2$ & $\phi_3$ & $\phi_4$ & $\chi_1$ & $\chi_2$ & $\chi_3$ \\
\hline
$SU(2)_L$ & 2 & 1 & 2 & 3 & 1 & 1 & 1 & 1 & 1 & 1 & 1 & 1 \\
$SU(2)_R$ & 1 & 2 & 2 & 1 & 3 & 1 & 1 & 1 & 1 & 1 & 1 & 1 \\
$U(1)_{B-L}$ & $-1$ & $-1$ & 0 & 2 & 2 & 0 & 0 & 0 & 0 & 0 & 0 & 0 \\
$S_4$ & $\mathbf{3}$ & $\mathbf{3}$ & $\mathbf{1}$ & $\mathbf{1}$ & $\mathbf{1}$ & $\mathbf{1}$ & $\mathbf{2}$ & $\mathbf{3'}$ & $\mathbf{3'}$ & $\mathbf{1}$ & $\mathbf{3'}$ & $\mathbf{3}$ \\
$Z_5$ & 1 & 4 & 2 & 0 & 0 & 2 & 2 & 2 & 2 & 4 & 4 & 4 \\
\hline\hline
\end{tabular}
\end{table}

\vspace{0.5cm}

\underline{Acknowledgments} \vspace{0.2cm}

This work was supported in part by the National Natural Science Foundation of China under Grant Nos.~12475112 and 12447101, Liaoning Revitalization Talents Program under Grant No.~XLYC2403152, and the Basic Research Business Fees for Universities in Liaoning Province under Grant No.~LJ212410165050.

\end{document}